\documentclass[twocolumn,showpacs,preprintnumbers,amsmath,prl,amssymb,superscriptaddress]{revtex4-1}

\pdfoutput=1

\usepackage[export]{adjustbox}
\usepackage{amsfonts}
\usepackage{amsmath}
\usepackage[makeroom]{cancel}
\usepackage[english]{babel}
\usepackage[T1]{fontenc}
\usepackage{times}
\usepackage{mathrsfs}
\usepackage{graphicx}
\usepackage{dcolumn}
\usepackage{bm}
\usepackage{wasysym}
\usepackage[colorlinks,bookmarks=true,citecolor=blue,linkcolor=red,urlcolor=blue]{hyperref}
\usepackage[tight, FIGTOPCAP, hang, raggedright, nooneline]{subfigure}
\usepackage{hyperref}
\usepackage{xcolor}
\usepackage{epsfig}
\usepackage{amssymb}
\usepackage{lipsum}
\usepackage{enumitem}

\newcommand{\subfigimg}[3][,]{%
  \setbox1=\hbox{\includegraphics[#1]{#3}}
  \leavevmode\rlap{\usebox1}
  \rlap{\hspace*{0pt}\raisebox{\dimexpr\ht1+0\baselineskip}{#2}}
  \phantom{\usebox1}
  }
  
 \definecolor{Green}{RGB}{80,182,0}

\usepackage{color}

\newcommand{\la}{\langle}
\newcommand{\ra}{\rangle}

\usepackage{epstopdf}

\begin{document}
\title{Dynamics of a Lattice Gauge Theory with Fermionic Matter -- \\ Minimal Quantum Simulator with Time-Dependent Impurities in Ultracold Gases
}
\author{Adam Smith}
\email{as2457@cam.ac.uk}
\affiliation{T.C.M. group, Cavendish Laboratory, J.~J.~Thomson Avenue, Cambridge, CB3 0HE, United Kingdom}
\author{Dmitry L.~Kovrizhin}
\affiliation{Rudolf Peierls Centre for Theoretical Physics, Parks road, Oxford, OX1 3PU, United Kingdom}
\affiliation{NRC Kurchatov institute, 1 Kurchatov square, 123182, Moscow, Russia}
\author{Roderich Moessner}
\affiliation{Max Planck Institute for the Physics of Complex Systems, N\"{o}thnitzer Stra{\ss}e 38, 01187 Dresden, Germany}
\author{Johannes Knolle}
\affiliation{Blackett Laboratory, Imperial College London, London SW7 2AZ, United Kingdom}
\date{\today}

\begin{abstract}
We propose a minimal model to study the real-time dynamics of a $\mathbb{Z}_2$ lattice gauge theory
coupled to fermionic matter in a cold atom quantum simulator setup. We show that
dynamical correlators of the gauge fields can be measured in experiments studying the
time-evolution of two pairs of impurities, and suggest the protocol for implementing
the model in cold atom experiments. Further, we discuss a number of unexpected features found
in the integrable limit of the model, as well as its extensions to a non-integrable case. A potential
experimental implementation of our model in the latter regime would allow one to simulate
strongly-interacting lattice gauge theories beyond current capabilities
of classical computers. 
\end{abstract}

\maketitle

\section{Introduction}

Due to remarkable experimental advances offering unprecedented control of isolated quantum systems, quantum simulators are becoming a reality, allowing one to test theoretical models of strongly-interacting quantum systems in the regimes beyond classical simulations. These recent experimental breakthroughs have been witnessed in a wide range of settings, including superconducting chips~\cite{Houck2012}, photonic quantum circuits~\cite{Aspuru2012}, and most notably in experiments with cold trapped ions~\cite{Martinez2016}, following early theoretical proposals, see e.g.~Ref.~\cite{Cirac1995}. Here we will focus on cold atom quantum simulator setups, such as recently used in the studies of many-body localization phenomena in two dimensions~\cite{Choi2016}.

An intriguing and natural application of quantum simulators is in the studies of lattice gauge theories (LGT)~\cite{Fradkin2013,Kogut1979,Montvay1994,Wiese2013,Zohar2016}, and in particular in testing their dynamical properties. Gauge theories play a central role in theoretical physics, from the standard model of fundamental particles to the low-energy descriptions of condensed matter systems. Important and well-known examples are those which are used in theoretical models of quantum chromodynamics (QCD), quantum electrodynamics (QED), as well as quantum spin liquids~\cite{Zhou2017}, and Kitaev's toric code~\cite{KitaevTQM}. While models appearing in condensed matter theory context are often lattice models, QCD and QED are continuum theories. However, they have also been modelled using LGT approaches~\cite{Chandrasekharan1997,Zache2018}. For example, a lattice version of the Schwinger model, which is a famous toy model of 1+1 dimensional QED~\cite{Montvay1994,Kogut1975}, has been recently simulated in cold ion trap experiments~\cite{Martinez2016}.

Digital (discrete time) quantum simulations of LGT have been so far restricted to one-dimensional systems because of requirements on high fidelity and on the large number of qubits in implementation of these simulations. Here, we propose a minimal setting for simulating the \textit{dynamics} of a LGT with fermions in 2+1 dimensions. We suggest that our model can be an ideal candidate for experimental implementations. First, we present a mapping of the model to free fermions (which can be studied exactly), and therefore the theory can be benchmarked against experiments. Via duality transformations we show that dynamical correlation functions of the gauge fields can be directly mapped to local impurity quenches in the free-fermion system. Second, even in the simple version of the model, where the free-fermion mapping holds, the model shows novel phenomenology of disorder-free localization~\cite{Smith2017,Smith2017_2,Smith2018}. Further, it can be tuned away from this `integrable' limit, where classical computation is no longer applicable. Third, measurements of correlation functions can be implemented using current technology in cold atomic gases~\cite{Choi2016}, and we propose a simple protocol based on Ramsey interferometry~\cite{Goold2011,Knap2012,Hangleiter2015,Streif2016} in a cold atom setting.

We note that there has been a large number of papers related to quantum simulations of gauge theories, and we refer the reader to Refs.~\cite{Wiese2013,Zohar2016,Dalmonte2016} for more detailed information.

\section{$\mathbb{Z}_2$ lattice gauge theory with fermionic matter}

One of the goals in the field of quantum simulation is to be able to test models of interacting quantum field theories, including  
the ones used in QCD, and QED. In this paper we suggest a lattice description of a version of a QED Hamiltonian with gauge fields coupled to fermions, and we focus on a 2+1D case, although the model can be studied in any dimensions. The continuum version of the model reads (here $\tilde{h},\tilde{K}$ are coupling constants, $m$ the fermion mass, $A$ the vector potential, and $E,B$ are the electric and magnetic field strengths)
\begin{equation} \label{H:QEDmod}
\begin{aligned}
H_{\text{cont}}=  \int d^2 x [-\frac{1}{2m} \bar \psi(p-A)^2\psi+\tilde{h} (\text{div}{E})^2 +\tilde{K} B^2].
\end{aligned}
\end{equation}
Note a non-standard $(\text{div} E)^2$, in other words the energy depends only on the divergence of the electric field, but not on the field strength, see also a discussion in \cite{Prosko2017}.

We discretise the model (\ref{H:QEDmod}) by placing fermions on the sites of a 2D square lattice and introduce a discrete $\mathbb{Z}_2$ vector potential and electric field on the links in the standard way, see e.g.~\cite{Prosko2017,Montvay1994,Wiese2013,Zohar2016}, and arrive at the discrete $\mathbb{Z}_2$ lattice gauge theory version of Eq.~\eqref{H:QEDmod},
\begin{equation} \label{H:lattspin}
H_{\text{lat}}  = - J\sum_{\la ij\ra} \hat{\sigma}^z_{ij} \hat{f}_i^{\dagger} \hat{f}_j-h\sum_i \hat A_i - K \sum_{p} \hat B_p,
\end{equation}
where $h,K,J$ are coupling constants, $\hat{f}^{\dagger}_i$ are spinless fermion creation operators on sites $i$, $\sigma^z_{ij}$ are the Pauli matrices defined on the links between neighbouring $i$ and $j$ sites. The star $\hat{A}_i$ and plaquette $\hat{B}_p$ operators (which are well-known from the context of the toric code), which live on the sites $i$ and plaquettes $p$ are correspondingly defined as
\begin{equation}
\hat{A}_i = \prod_{j : \la i j \ra} \hat{\sigma}^x_{ij}, \qquad \hat{B}_p = \prod_{\text{plaquette }p} \hat{\sigma}^z.
\end{equation}
The model Eq.~\eqref{H:lattspin} is a natural two-dimensional generalization of the model studied by the authors in Ref.~\cite{Smith2018} in the context of disorder-free localization. The latter localization mechanism was later studied in the context of the 1D $U(1)$ lattice Schwinger model in Ref.~\cite{Brenes2018}. In this work we focus on dynamics of the gauge fields in the 2D case. One of the central results of this work is a protocol for quantum simulation of time-dependent gauge-field correlators after a quantum quench. We are interested in measuring the connected spin correlators
\begin{equation}\label{eq: spin correlator}
\la \hat{\sigma}^z_{jk}(t) \hat{\sigma}^z_{lm}(t) \ra_c = \la \hat{\sigma}^z_{jk}(t) \hat{\sigma}^z_{lm}(t) \ra -  \la \hat{\sigma}^z_{jk}(t)\ra \la \hat{\sigma}^z_{lm}(t) \ra.
\end{equation}

\section{Duality Mapping to Free Fermions and Dynamical Correlation Functions}

The model in Eq.~\eqref{H:lattspin} is the Kitaev Toric code with the additional term describing dynamics of free spinless fermions coupled to gauge fields via minimal coupling. Below we consider a quantum quench from an initial state $|\Psi\ra$ which is invariant under the application of all plaquette operators, i.e. $\hat{B}_p |\Psi\ra = |\Psi\ra$. Since these operators are conserved under dynamics we consider only the sector with eigenvalues $B_p = 1$ for every $p$.
More general quenches can be studied in a similar way, where one will have to take into account the disconnected sectors labelled by $B_p=\pm 1$, see Refs.~\cite{Smith2018,Prosko2017} for more details. Up to a constant, the Hamiltonian of Eq.~\eqref{H:lattspin} for a fixed sector takes the following form
\begin{equation}\label{eq: original H}
\hat{H} =   - J \sum_{\la i j \ra} \hat{\sigma}^z_{ij} \hat{f}^\dag_i \hat{f}_j -h \sum_i \hat{A}_i .
\end{equation}

In order to bring the model into a solvable form we introduce a transformation of degrees of freedom. First, we perform a standard duality transformation for the spin degrees of freedom, which is well-known in the context of the Ising model,
\begin{equation}
\hat{\tau}^z_j = \hat{A}_j, \qquad \hat{\sigma}^z_{jk} = \hat{\tau}^x_j\hat{\tau}^x_k,
\end{equation}
where $\hat{\tau}^x_{i},\hat{\tau}^z_{i}$ are the Pauli matrices defined on the sites $i$ of the lattice.
We can now identify conserved charges $\hat{q}_j = \hat{\tau}^z_j e^{i\pi\hat{f}^\dag_j\hat{f}^{}_j}$ with the eigenvalues $\pm1$, and introduce the following transformations for the fermion operators $\hat{c}_j = \hat{\tau}^x_j \hat{f}_j$. The $\hat{\tau}$ and the $\hat{c}$ operators obey standard commutation relations. The Hamiltonian commutes with the operators of conserved charges $[\hat{H},\hat{q}_j] =0$, and the charges commute between themselves $[\hat{q}_j, \hat{q}_k] = 0$. In terms of these $\hat{c}$ and $\hat{\tau}$ operators, the Hamiltonian assumes the form (see~\cite{Smith2017,Smith2017_2,Smith2018,Prosko2017})
\begin{equation}\label{eq: free fermions}
\hat{H} = -J \sum_{\la j k \ra} \hat{c}^\dag_j \hat{c}_k + 2h \sum_j \hat{q}_j (\hat{c}^\dag_j \hat{c}_j - 1/2).
\end{equation}
The model in Eq.~(\ref{eq: free fermions}) is a free-fermion model, and can be studied using standard techniques, which we outline below on the example of a quantum quench problem. Note that the conserved charges in this case will be defined by the initial state of the system.

\subsection{Quantum Quench Protocol}

Below we focus on a quantum quench problem where the spins and the fermions are prepared in some initial state, and we will calculate the dynamics of spin-correlation functions at time $t$ after the quench. An interesting, and simple initial state is given by a tensor product of the spins polarized along the $z$-axis and fermions in a Slater determinant state at half-filling $|\Psi\ra = |\!\uparrow\uparrow \cdots\ra\otimes|\psi\ra$. The fermion Slater determinant describes a Fermi-sea for the Hamiltonian $\hat{H}_\text{FS} = -\sum_{\la i j \ra} \hat{f}^\dag_i \hat{f}_j$. In terms of the charges $\hat{q}$ and the fermions $\hat{c}$ these initial states take the form
\begin{equation}
|\Psi\ra = \frac{1}{\sqrt{2^{N-1}}} \sideset{}{'}\sum_{\{q_i\}=\pm 1} |q_1 q_2 \cdots q_N \ra\otimes |\psi\ra,
\end{equation}
where the primed sum is over all charge configurations $\{q_i\}$ such that $\prod_\text{all}\hat{q}_j = (-1)^{N_f}$, where $N_f = \sum_i\hat{f}^\dag_i\hat{f}^{}_i$ is the fermion filling, see Refs.~\cite{Smith2017,Smith2017_2,Smith2018} for more details.

\begin{figure}[t!]
\includegraphics[width=.49\textwidth,valign=b]{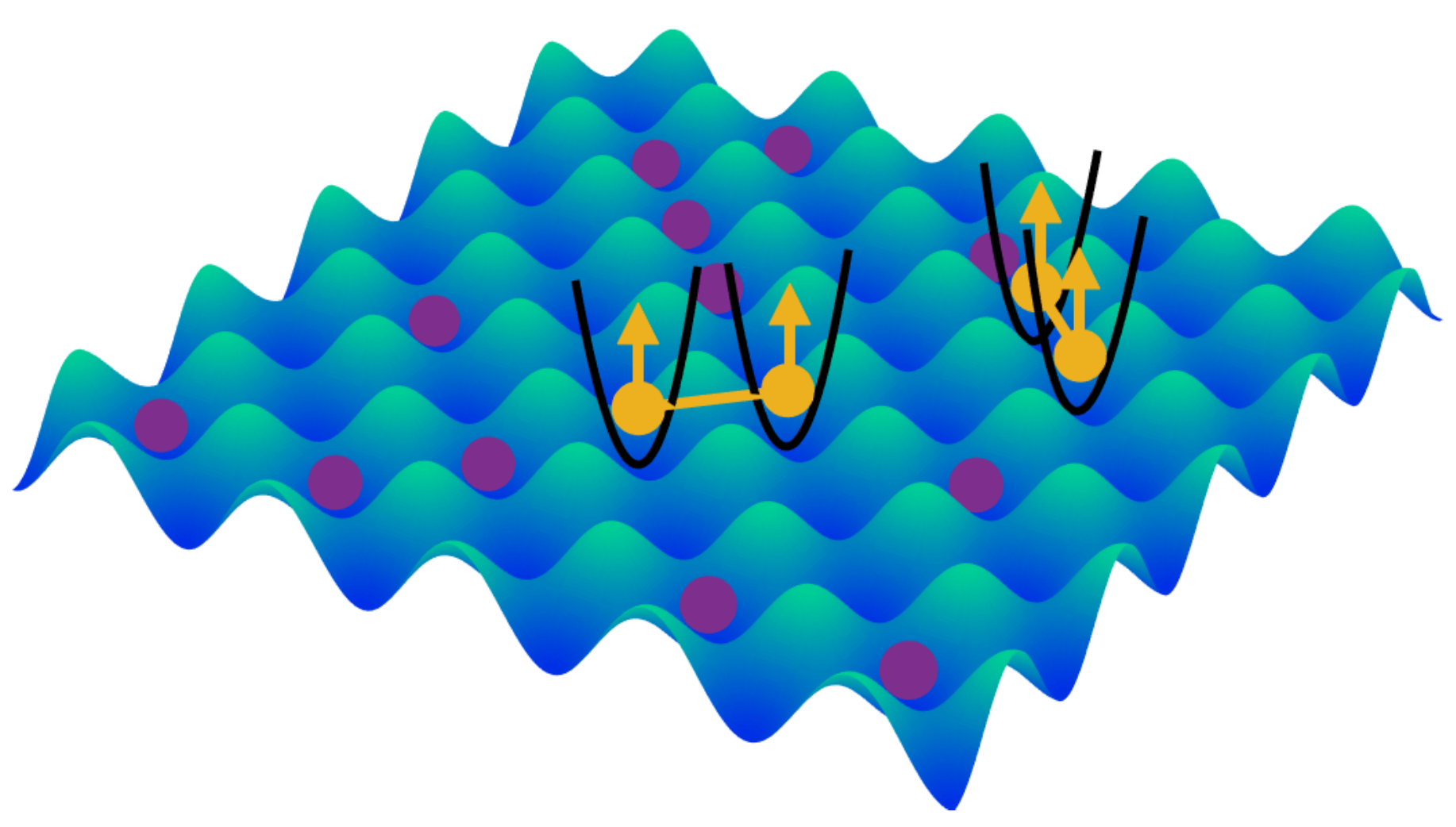}
\caption{Schematic picture of the proposed experimental setup. Fermions (purple balls) are confined in an optical lattice (blue surface) with nearest-neighbour hopping between lattice sites described by the Hamiltonian Eq.~\eqref{eq: free fermions}. The fermions experience a binary potential set by charges $\{q_i\}= \pm 1$, except at particular sites where we put spin-1/2 impurities (yellow). Impurity spins are localized in separately controlled much deeper wells, and generate an external potential on these sites, with the sign of the potential for spin up/down impurities being positive/negative correspondingly. To calculate  spin correlators, corresponding to gauge-field correlators, one has to control four impurity spins, with pairs of spins located on the sites sharing the bond associated with the gauge spins. The correlators are then calculated using the Loschmidt echo protocol, defined in the main text, which involves a $\pi/2$-rotation and measurement of the impurity spins.}\label{fig: Optical Lattice}
\end{figure}

Let us now discuss the calculation of the dynamical correlators. The simplest component of the connected spin-spin correlator is the average the local magnetisation
\begin{equation}
\la \hat{\sigma}^z_{jk}(t)  \ra = \la \Psi | e^{i\hat{H}t} \hat{\tau}^x_j \hat{\tau}^x_k e^{-i\hat{H}t} |\Psi\ra.
\end{equation}
In order to simplify notation we introduce a Hamiltonian for a fixed configuration of charges $\{q_i\} = \pm1$,
\begin{equation}
\hat{H}(q) = -J \sum_{\la j k \ra} \hat{c}^\dag_j \hat{c}_k + 2h \sum_j q_j (\hat{c}^\dag_j \hat{c}_j - 1/2),
\end{equation}
together with the short hand notation $\hat{H}_{jk}(q)$ for the Hamiltonian having the sign of $q_j,q_k$ flipped relative to $\hat{H}(q)$, then the magnetization at time $t$ after a quench is given by the following correlator
\begin{equation}\label{eq: S magnetization}
\la \hat{\sigma}^z_{jk}(t)  \ra = \frac{1}{2^{N-1}} \sideset{}{'}\sum_{\{q_i\}=\pm1} \la \psi | e^{i\hat{H}_{jk}(q)t} e^{-i\hat{H}(q) t} |\psi \ra.
\end{equation}
This has the form of binary disorder-averaged Loschmidt echo, where the charges $\hat{q}$ at sites $j,k$ having opposite signs in the forward and backward time-evolution. Repeating the same arguments we obtain the expression for the connected spin-correlator at time $t$ after the quench
\begin{equation}\label{eq: SS correlator}
\la \hat{\sigma}^z_{jk}(t) \hat{\sigma}^z_{lm}(t)  \ra = \frac{1}{2^{N-1}} \sideset{}{'}\sum_{\{q_i\}=\pm1} \la \psi | e^{i\hat{H}_{jklm}(q)t} e^{-i\hat{H}(q) t} |\psi \ra,
\end{equation}
where one has to exchange signs of four charges at sites denoted by $jklm$ between forward and backward evolution. Eq.~\eqref{eq: SS correlator} is one of the central new results of this work. The dynamical correlation function of the gauge field directly correspond to a local quantum quench of a (free) fermionic Hamiltonian. In the following section we discuss how the latter can be efficiently simulated in a cold atom setup. 

The Loschmidt echo appearing in these expressions can be efficiently computed numerically using fermion determinants. For example, for the expression Eq.~\eqref{eq: S magnetization} these take the form
\begin{equation}
\la \psi | e^{i\hat{H}_{jk}(q) t} e^{-i\hat{H}(q) t} |\psi \ra = \det[V^\dag U_{jk}^\dag(q) U(q) V],
\end{equation}
where $U(q) = e^{-iH(q) t}$ is the exponential of the matrix $H(q)$, and similarly for $U_{jk}(q)$ and $H_{jk}(q)$, and $V$ is a rectangular matrix which has as its columns the $N/2$ filled eigenvectors of the Hamiltonian $\hat{H}_\text{FS} = -\sum_{\la i j \ra} \hat{f}^\dag_i \hat{f}_j$. The determinants for the two-point correlators~\eqref{eq: SS correlator} take a similar form but with $H_{jklm}(q)$ instead of $H_{jk}(q)$.

\section{Quantum Simulation and Experimental Setup}

A schematic picture for the experimental setup that we propose is shown in Fig.~\ref{fig: Optical Lattice}. We consider a square optical lattice with fermions prepared in a Fermi-sea at half-filling. At time $t=0$ we abruptly turn-on a binary disorder potential -- similar to the quantum gas microscope set up used in Ref.~\cite{Choi2016} -- and we also add two or four impurity spins which control the potential in order to obtain the Loschmidt echo~\cite{Knap2012,Hangleiter2015,Streif2016}. The impurities should be localized by an external trapping potential, and one has to chose these impurity spins such that the lattice fermions should interact strongly with one of the two spin states $|\!\uparrow_j\ra$ and weakly with the other $|\!\downarrow_j\ra$, effectively turning on/off the potential for the fermions on the impurity site.

\begin{figure*}[tb]
\centering
\subfigimg[width=.2\textwidth,valign=t]{\hspace*{8pt} \textbf{(a) Horizontal $h = 0.7J$}}{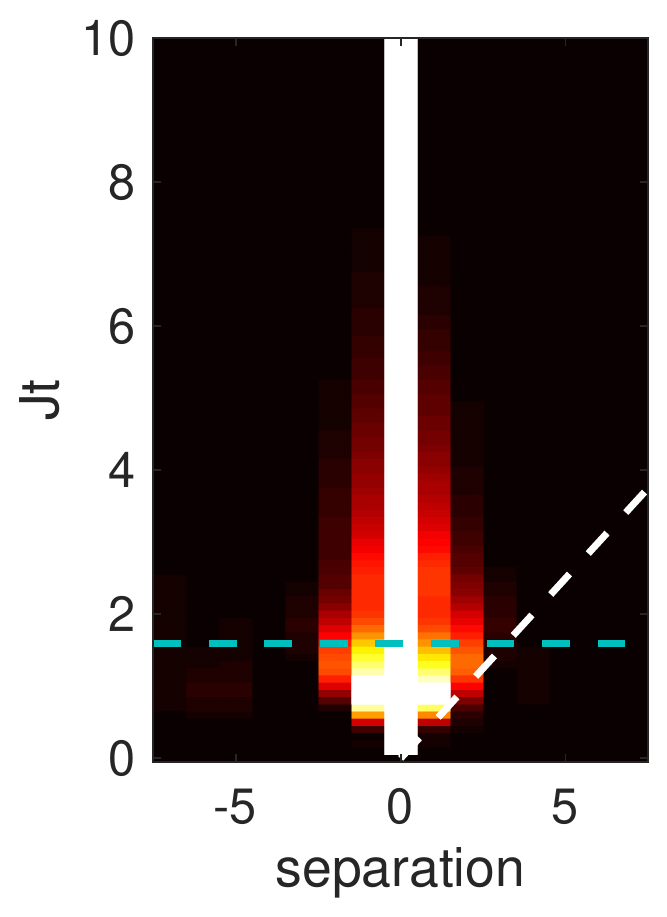}\quad
\subfigimg[width=.2\textwidth,valign=t]{\hspace*{8pt} \textbf{(b) Diagonal $h = 0.7J$}}{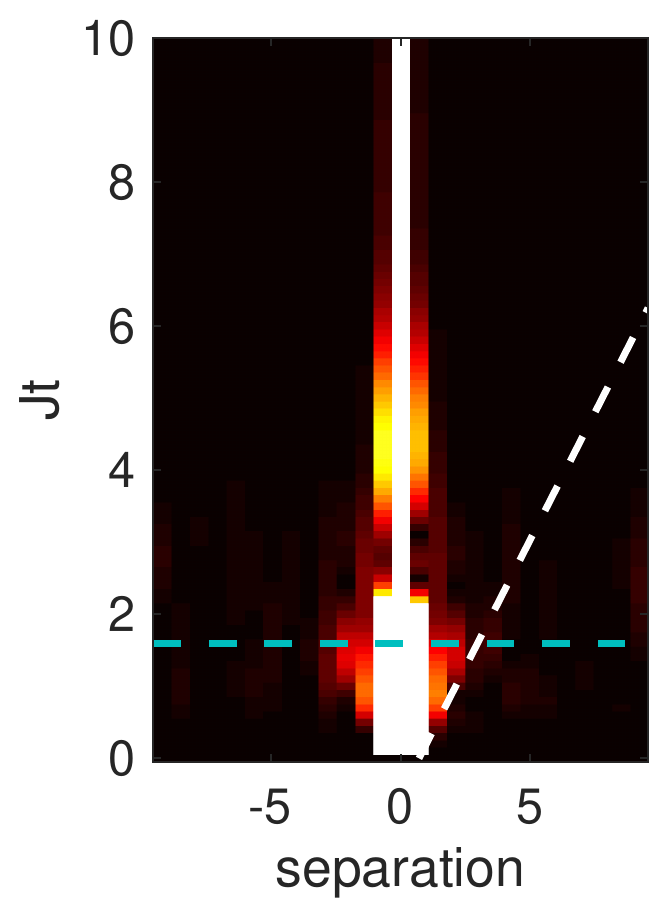}\quad
\subfigimg[width=.2\textwidth,valign=t]{\hspace*{8pt} \textbf{(c) Horizontal $h = 2J$}}{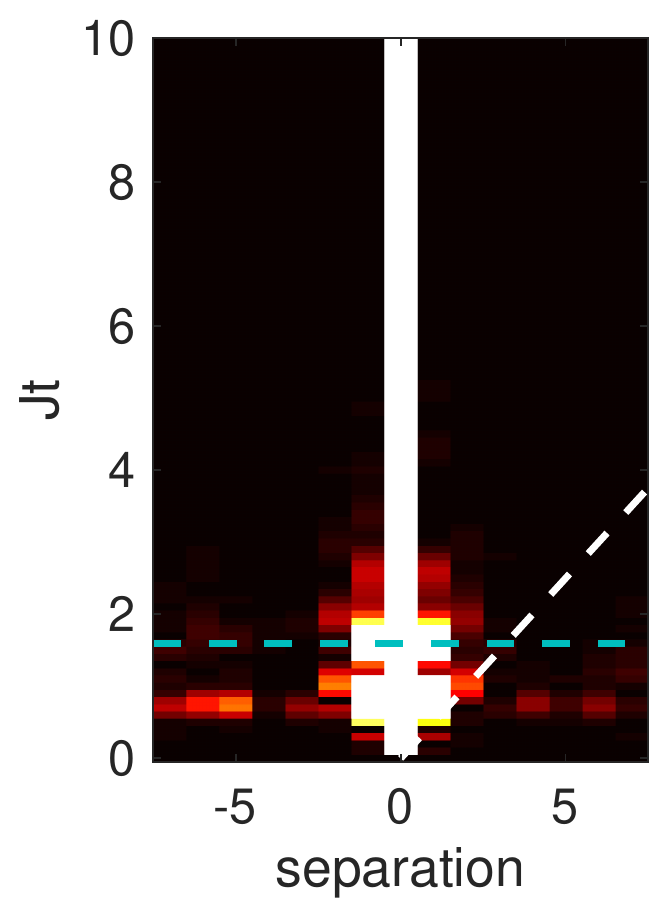}\quad
\subfigimg[width=.2\textwidth,valign=t]{\hspace*{8pt} \textbf{(d) Diagonal $h = 2J$}}{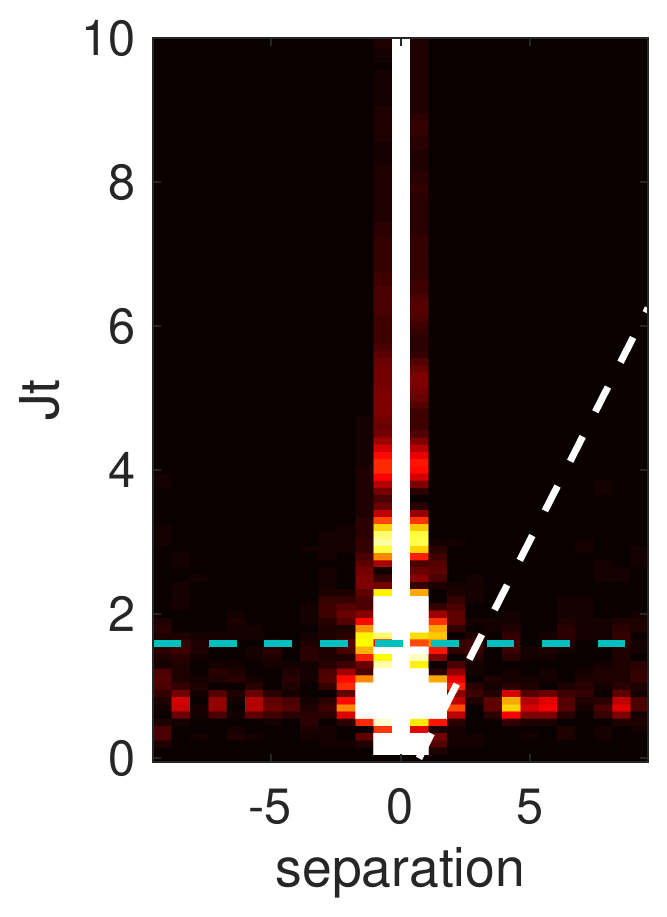}\qquad
\subfigimg[width=.057\textwidth,valign=t]{\hspace*{0pt} \textbf{}}{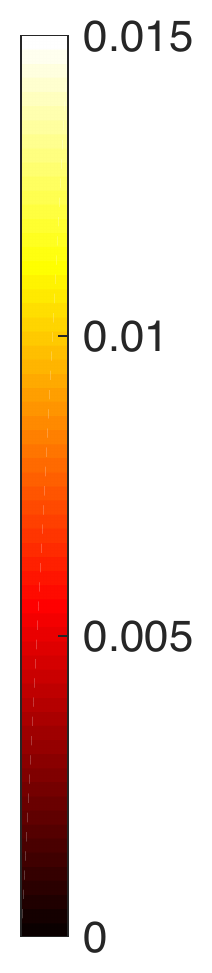}
\caption{Absolute value of the connected spin-spin correlator $|\la \hat{\sigma}^z_b(t) \hat{\sigma}^z_0(t) \ra_c| $ for different spatial cuts across the system as a function of time and separation between the spin $\hat{\sigma}^z_b$ and the spin $\hat{\sigma}^z_0$. Results are shown for two different values of $h/J = 0.7,2$. The horizontal and diagonal cuts across the lattice are shown in Figs.~\ref{fig: slices}. The blue dashed line indicates the time slice that is shown in Fig.~\ref{fig: slices}. The white dashed line corresponds to the the light-cone propagation with velocity $v = 2J$.}\label{fig: cuts}
\end{figure*}

As shown above, the correlators that we are interested in can be mapped to the measurement of the Loschmidt echo. We prescribe an implementation of this measurement using Ramsey interferometry, which we describe below. Further details on experimental implementations can be found in~\cite{Goold2011,Knap2012,Hangleiter2015,Streif2016}. The impurity spins in the up state interact with the fermions via a local interaction, while the spins in down states and the fermions are decoupled.  Now we introduce a composite two state system, which we will call a control spin. In the measurement of the average local magnetisation the  control spin affects two nearest-neighbour charges, i.e. $|\!\Downarrow\,\ra \leftrightarrow |\!\downarrow_j \downarrow_k\ra$, $|\!\Uparrow\,\ra \leftrightarrow |\!\uparrow_j \uparrow_k\ra$, and in the measurement of two point correlators we need two pairs of impurity spins (as shown in Fig.~\ref{fig: Optical Lattice}), i.e., $|\!\Downarrow\,\ra \leftrightarrow |\!\downarrow_j \downarrow_k\downarrow_l \downarrow_m\ra$, $|\!\Uparrow\,\ra \leftrightarrow |\!\uparrow_j \uparrow_k\uparrow_l \uparrow_m\ra$.
Using the average local magnetisation as a specific example, the measurement protocol is the following:
\begin{description}[align=left]
\item [Initialise] The system is initialised in the state $|\psi\ra$ with fermions forming a half-filled Fermi-sea for the Hamiltonian  $\hat{H}_\text{FS} = -\sum_{\la i j \ra} \hat{f}^\dag_i \hat{f}_j$, and the control spin is prepared in the state $|\!\Downarrow\,\ra$.
\item [$\pi/2$ pulse] We perform a $\pi/2$ pulse on the control spin such the state of the fermions and control spins becomes $|\Psi\ra =\frac{|\Uparrow\ra + |\Downarrow\ra}{\sqrt{2}}\otimes |\psi\ra$.
\item [Evolve] We let the system evolve, so that its state at time $t$ is given by
\begin{equation}
\qquad|\Psi(t) \ra = \frac{1}{\sqrt{2}}[e^{-i \hat{H}(q) t} |\!\Downarrow\,\ra\otimes|\psi\ra + e^{-i \hat{H}_{jk}(q) t} |\!\Uparrow\,\ra\otimes |\psi\ra ].
\end{equation}
\item [$\pi/2$ pulse] At time $t$ we perform the reverse $\pi/2$ pulse on the control spin so that we can perform a measurement in the original basis of impurity spins.
\item [Measure] We measure the $z$-component of the control spin, which gives
\begin{equation}
\qquad\la \hat{S}^z \ra = \text{Re}\la\psi | e^{i \hat{H}_{jk}(q) t} e^{-i \hat{H}(q) t} | \psi \ra.
\end{equation}
\end{description}
This procedure must be repeated for different disorder realisations and the result averaged over these measurements. In the case of the two-point correlator the control spin corresponds to two pairs of impurities which amounts to replacing $\hat{H}_{jk}(q)$ by $\hat{H}_{jklm}(q)$. From our calculations we find that it is sufficient to average over a small subset of random binary disorder realisations. Because these spin correlators must be real, this procedure amounts precisely to the calculations we wish to perform in Eq.~\eqref{eq: spin correlator}.

\section{Numerical Results}

\begin{figure}[b]
\centering
\subfigimg[width=.2\textwidth,valign=t]{\hspace*{-9pt} \textbf{(a) Slice $h = 0.7J$}}{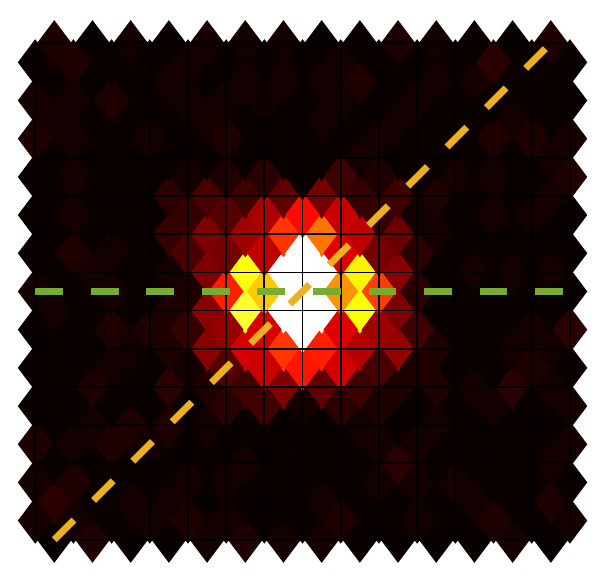}\;\;
\subfigimg[width=.2\textwidth,valign=t]{\hspace*{-9pt} \textbf{(b) Slice $h = 2J$}}{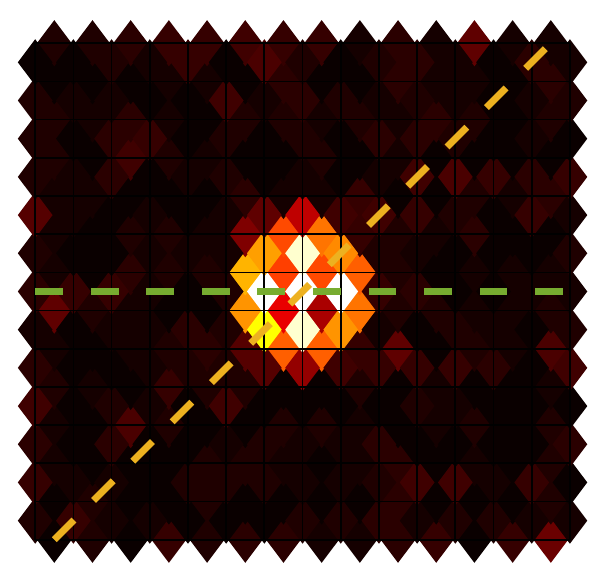}\;\;
\subfigimg[width=.051\textwidth,valign=t]{\hspace*{0pt} \textbf{}}{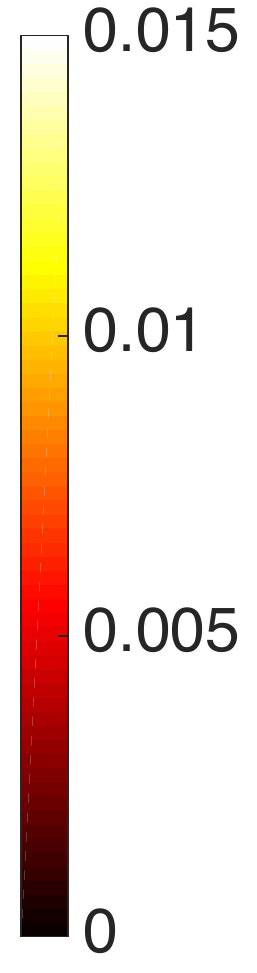}
\caption{Spatially resolved absolute value of the connected spin-spin correlator $|\la \hat{\sigma}^z_b(t) \hat{\sigma}^z_0(t) \ra_c| $. The spin $\hat{\sigma}^z_0$ resides on the central bond and $\hat{\sigma}^z_b$ is taken on the other bonds of the $15\times14$ lattice for $Jt = 1.7$ and $h/J = 0.7,2$. Superimposed is the lattice (black) where fermions reside. The dynamics along the horizontal and diagonal cuts indicated by dashed lines are shown in Fig.~\ref{fig: cuts}.}\label{fig: slices}
\end{figure}

Here we present some numerical results for a $15\times14$ square lattice for $h/J = 0.7,2$ where we used averaging over $1000$ disorder realisations. In Fig.~\ref{fig: cuts} we show the time dependence of the connected two-point spin correlator~\eqref{eq: spin correlator} as a function of separation between two spins along the horizontal and diagonal cuts indicated by dashed lines in Fig.~\ref{fig: slices}. Two main features common to all four figures are the linear light-cone spreading and the eventual decay of all spatial correlations. 

\begin{figure}[t!]
\includegraphics[width=.42\textwidth,valign=b]{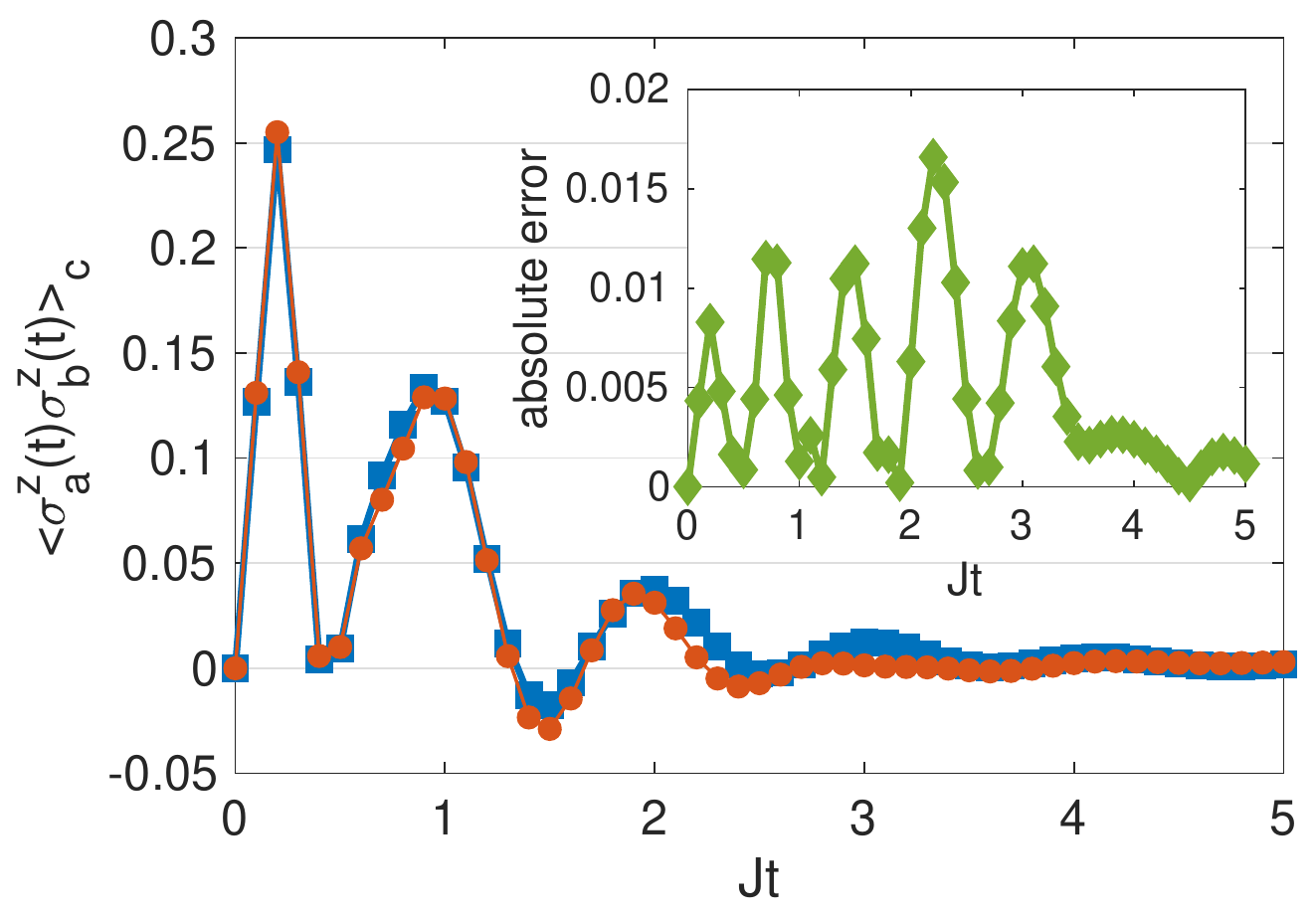}
\caption{Comparison of the nearest-neighbour correlator $\la \hat{\sigma}^z_b(t) \hat{\sigma}^z_0(t) \ra_c$ along the diagonal cut indicated in Fig.~\ref{fig: slices} for different number of disorder realisations, $h/J = 2$. (Blue line) 1000 disorder realisations, (Red line) 50 disorder realisations. (Inset) Absolute value of the difference between the results for 50 and 1000 disorder realizations.}\label{fig: low samples}
\end{figure}

The spreading of correlations is linear in all cases and has velocity $v = 2J$, which is the maximal group velocity of the fermions. This light-cone regime is short-lived due to the overall decay of the correlations. A notable difference between the horizontal and diagonal cuts is that the correlations between the neighbouring spins along the diagonal grow immediately, leading to a slightly offset light-cone. This can be appreciated by looking at Fig.~\ref{fig: slices} where one can see that these two spins belong to the same star operator and thus the correlations begin to grow after the quench with the rate set by $h/J$, as shown in Fig.~\ref{fig: low samples}. Difference appears when we increase $h$, where we see that the peaks in the correlations become much sharper along the light-cone, which is followed by decaying oscillations. The spatial pattern of correlations also changes as shown for a time slice $Jt = 1.7$ in  Fig.~\ref{fig: slices}. The extent of the spreading is bigger for smaller $h/J$. Further, for small values of $h/J$ one can notice the asymmetry due to the fact that the central bond is vertical and we don't have 90 degree rotational symmetry, whereas for $h=2J$ this asymmetry seems to be smaller.

While for an exact simulation of the gauge field we would need to average over all possible configurations of the potential generated by the conserved charges, in order to obtain accurate results we require only a tiny fraction of configurations. For $h/J=0.7$, we see that the results have the correct symmetry and there is very little noise (due to finite number of disorder realizations). On the other hand, for $h/J=2$ there seem to be more-pronounced non-physical correlations, most notably appearing as a stripe in Figs.~\ref{fig: cuts}(c-d) at around $Jt=1$, which is responsible for partially obscuring the light-cone. We can also see a faint non-uniform random background in Fig.~\ref{fig: slices}(b). In order to minimise these artefacts one has to use a larger number of disorder realisations. Despite these issues, qualitatively good results can be obtained with as few as 50 disorder configurations as shown in Fig.~\ref{fig: low samples} for the nearest-neighbour correlators along diagonal cut. With a very small number of samples one can extract a number of qualitative features such as sharp growth of correlations and the decaying oscillations.

\section{Discussion}

We have presented a minimal 2D model of a $\mathbb{Z}_2$ lattice gauge theory coupled to fermionic matter and outlined an experimental protocol for measuring time-dependent gauge field correlation functions. We believe that this protocol should be accessible with current experimental technology. In experiments with cold atoms in optical lattices, such as discussed in Ref.~\cite{Choi2016}, relatively large 2D systems have already been simulated, and the protocol that we propose is designed to require minimal extra complications. Using a duality mapping to free-fermions we are able to perform efficient numerical computations which should allow one to quantitatively compare theory and experiment. Some clear features of dynamical correlations, such as the light-cone spreading and the decay of the correlations at long times should be directly accessible to experimental measurements. 

We anticipate a number of possible practical issues regarding the difference between the experimental setup and our model, which should not be very difficult to account for. These may include an actual realization of the binary potential, the shape of a trapping potential for fermions, or realization of the coupling to impurity spins. All of these features can be easily implemented in our theoretical calculations. Further, one can access relatively large system sizes in numerical calculations, which would allow one to look at the scaling in the system size and the number of disorder realisations.

While the dynamics of the model in Eq.~\eqref{eq: original H} can be efficiently simulated on a classical computer, the model offers a number of generalisations which turn it into a strongly-interacting quantum model. Specifically, one can add to the Hamiltonian additional terms which will be still commuting with the charges. This means that the mapping and experimental protocol presented above are still valid. For example, one can add nearest-neighbour density interactions between fermions $\sum_{\la j k \ra} \hat{n}_j \hat{n}_k$ which have the same form in the original and the dual representation. In the 1D case the model maps onto an XXZ spin chain with binary disorder potential, realizing the minimal model of MBL behaviour. While there is no such mapping in 2D, the physics of MBL may be relevant in this case as well. One can also generalise our model using spin-1/2 fermions. In this case our model can be identified with the slave-spin description of the Hubbard model~\cite{Ruegg2010,Zitko2015}, but without Gauss' law imposed. With the addition of interactions, classical computations can only access very small system sizes in two dimensions (with some exceptions). However, in experiment, these generalisations are not expected to present significant extra difficulties. 

To conclude, our model offers an ideal setting for simulating LGT dynamics in two-dimensions in the regimes beyond classical computation capabilities, with the important feature of a well defined free-fermion limit which provides theoretical results in a non-trivial setting.

\section*{Acknowledgements} We are grateful to Ulrich Schneider for enlightening and encouraging suggestions. A.S.~acknowledges EPSRC for studentship funding under Grant No.~EP/M508007/1. The work of D.K.~was supported by EPSRC Grant No.~EP/M007928/2.  R.M.\ was in part supported by DFG under grant SFB 1143.

\end{document}